# Running Coupling in SU(3) Yang-Mills Theory

Ulli Wolff [a] *

[a]CERN TH-Division, CH-1211 Geneva 23, Switzerland

We report about our ongoing computation of running coupling constants in asymptotically free theories using the recursive finite size scaling technique. The latest results for the SU(3) Yang-Mills theory are presented.

## 1. INTRODUCTION

In nonperturbative evaluations of asymptotically free theories all input parameters can be fixed by matching low energy physical quantities, like the mass spectrum in QCD, with experiment. A physical renormalized coupling constant at high energy is then a *computable* number. If such a computation is achieved with sufficient precision for some suitable coupling $\alpha(q)$, then the connection with the whole perturbative sector of the theory can be made by expanding in powers of $\alpha(q)$. One has then established a link between the high and low energy sectors. Of course, with only a few perturbative orders available, it is essential to use a non-pathological $\alpha(q)$ and to control its values over an adequate range of high energies. It is important to understand that such a computation of absolute numbers for $\alpha(q)$ — even at large $q$ where it *evolves* perturbatively with $q$ — is a nonperturbative problem. As we fix parameters at low energy, we have to specify $q$ as a dimensionless multiple of some low energy quantity like a mass $m$ or the string tension in the quenched theory. The difficulty of such a calculation is mainly due to the fact that the ratio of scales $q/m$ should be large, at least $O(10)$. In addition, at least with conventional approaches (see below), the lattice spacing $a$ and system size $L$ have to be remote of either $m$ or $q$.

A straightforward idea to compute the coupling in pure gauge theory is to proceed via the static quark-antiquark force $F(r)$. While it saturates to the string tension (the only free parameter) at large distance, $r^2 F(r)$ can also be used as a physical running coupling constant at small separation $r$. This requires control of the force from short to long distance in one simulation with $a \ll r \ll L$ holding for the whole range of physical $r$ involved. With $L/a$ always limited to feasible lattice sizes like 32 or 48, compromises on the above conditions have to be accepted, and it is hardly possible to vary all scale ratios significantly to check for the stability of the results. State of the art calculations along these lines are reported in [1,2]. It has to be noted that the highest physical energies $r^{-1}$ that can be reached here are below about 2 GeV, if one only stays a factor 2...3 away from the cutoff energy. Cutoff effects are corrected semi-empirically using the lattice Coulomb propagator. While these are difficult and careful simulations, we find it somewhat hard to assess the systematic errors in a completely convincing fashion.

An alternative attempt to derive the coupling in QCD has been pioneered by the Fermilab group [3]. Here, in a quenched simulation, the spin averaged 1P-1S charmonium splitting is determined on a physically large lattice. Although this is a nice experimentally known scale with little sensitivity to the quark masses, also other masses could in principle be used here to set the scale. The point relevant in the present context is, that they extract from such a simulation the bare lattice coupling $g_0$ together with the corresponding lattice spacing $a$ in GeV. A perturbative method is then used to relate $g_0$ to a physical coupling at a scale of the order of the cutoff. The scale problem is clearly alleviated in comparison to the quark force method, as effectively the cutoff $a^{-1}$ is identified with the high energy physical scale.

---

*Address after april 1, 1994: FB Physik, Humboldt Universität, Invalidenstr. 110, D-10099 Berlin, Germany

The problem is now to convert reliably from the non-universal unphysical lattice-coupling $g_0$ to a physical continuum one like $\alpha_s(q = \pi/a)$ with one non-trivial perturbative expansion coefficient (presently) known. In [4] it is shown that a large number of numerically known quantities, Wilson loops for instance, are well approximated by the so-called improved perturbation theory. This is used in [3], with the average plaquette as the key input parameter to implement improved perturbation theory. Again, one may feel that there could be a problem with estimating the error in the conversion.

In the following section I recall the recursive finite size method and present the latest results obtained.

## 2. RUNNING COUPLING BY RECURSIVE FINITE SIZE SCALING

The ideas and techniques behind this approach were presented in detail in Lüscher's talk in Amsterdam [5], and we content ourselves here with a brief reminder. The freedom which coupling to compute is exploited by choosing $\alpha(q) = \bar{g}^2(L)/4\pi, q = 1/L$, a coupling that runs with the boxsize $L$. It's definition is based on an abelian background field that is induced through nontrivial boundary conditions [6]. A physical change in the free energy with a variation of the field leads to the coupling $\bar{g}^2(L)$. The definition for general $L$ is independent of perturbation theory, and at small $L$ we expect $\bar{g}$ to be a smooth perturbative coupling. As in the Fermilab approach one of the scales unavoidable in numerical treatments is used for physics here to eliminate one large scale ratio. We think, however, that $L$ has the advantage over $a$ of finite size effects being universal.

A further characteristic of the method is to break the problem into several steps. In each step of a series of simulations we numerically answer the question: Given $\bar{g}^2(L) = u$, with $u$ some number like 1.234 implicitly defining the scale $L$, what is $\bar{g}^2(2L)$? This result equals the value of the step scaling function,

$$\sigma(2, u) = \lim_{a \to 0} \Sigma(2, u, a/L), \qquad (1)$$

where $\Sigma$ refers to a sequence of finite lattice realizations with decreasing $a$ that are to be extrapolated to the universal continuum function $\sigma$. The values, for which $\sigma$ is determined are tuned such that we can iterate $\sigma(2, \sigma(2, ...))$ and thus recursively get $\sigma$ for scale factors larger than two. In each individual step we only have to deal with scales $a$ and $L$, and we use all the available $L/a$ to determine and extrapolate the cutoff dependence. All our results for $\Sigma$ in the pure SU(3) gauge theory [7] are shown in Fig. 1.

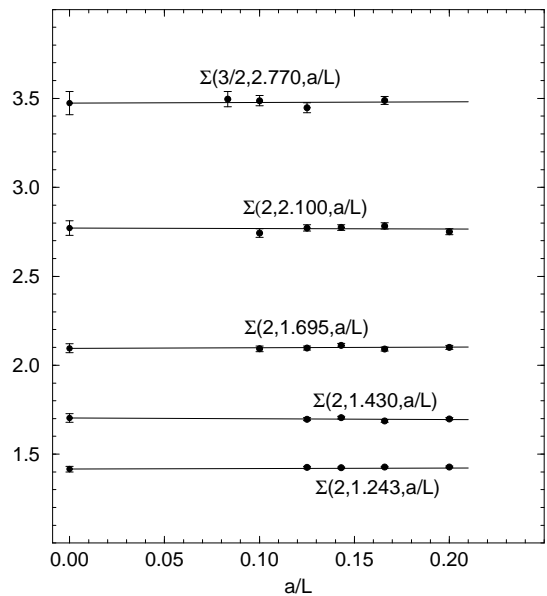

Figure 1. Extrapolation of the step scaling function to the continuum.

Obviously there is very little cutoff dependence, which is probably due to our use of a 1-loop $O(a)$ improved Wilson action. The extrapolation to the continuum, which on theoretical grounds is expected to be roughly linear in $a/L$ for our background field, presents no problem. Iterating the extrapolated values we gain information about $\sigma(s, 1.243)$ with $s \approx 2, 4, 8, 16, 24$ [In the last step we used a factor 3/2 instead of 2, and the precise $s$ values differ slightly from these integers as there are small mismatches in the iteration, which are easily taken into account. Moreover, all numbers have statistical errors. We omit these details in this text, they can be found in [7] and are included in all figures]. The largest coupling occurring is $\bar{g}^2(s_{max}L) = 3.48$ with $s_{max} \approx 24$ and $\bar{g}^2(L) = 1.24$. We expect $s_{max}L = L_{max}$ to be in the range of nonperturbative scales. Its precise relation to such a scale has to be determined in the last step.

## 3. PHYSICAL SCALE AND COUPLING

Rather than setting the scale through the string tension, we use the recently proposed [8] unit of length $r_0$ defined from the interquark force by

$$r_0^2 F(r_0) = 1.65. \qquad (2)$$

If the force is identified with the one used in charmonium potential calculations, then $r_0 \approx 0.5$ fm is suggested. Advantages of $r_0$ over the string tension are discussed in [8]. On the basis of potential data from the literature, $L_{max}/r_0$ assumes the value $0.674(50)$ after extrapolating $a \to 0$ [7]. Our results are summarized in Fig. 2. The pertur-bative evolutions start from the smallest coupling and the fit is made with an effective 3-loop term. The error in the overall scale, which corresponds to a horizontal shift of all points and curves in the semi-logarithmic plot, is not included in Fig. 2.

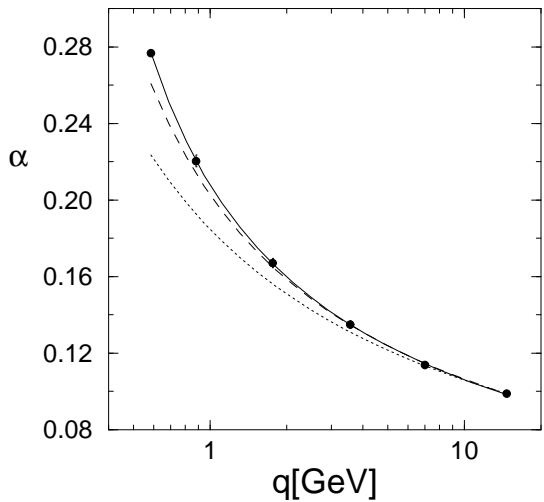

Figure 2. Values of the running coupling together with perturbative evolution to 1-loop (dotted) and 2-loop (dashed), and a fit (solid).

At this point we have learned that $\bar{g}^2(L = 0.027(3)r_0) = 1.243$. It remains to pass from our finite volume coupling to $\alpha_s$ which by convention refers to the $\overline{MS}$-scheme. At $q = L^{-1}$ we should be well in the perturbative regime and can use the 1-loop formula [7]

$$\alpha_{\overline{MS}}(q) = \alpha(q) + 1.2556\, \alpha(q)^2 + O(\alpha(q)^3). \qquad (3)$$

The final result at the highest energy is now

$$\alpha_{\overline{MS}}(q) = 0.1108(23)(10) \text{ at } q = 37\, r_0^{-1}, \qquad (4)$$

which corresponds to about 15 GeV. The second error is the estimated effect of missing higher orders in (3) and should be eventually eliminated by a 2-loop calculation, while the first one combines all other errors. As is obvious from Fig. 2, apart from computing (4) we have demonstrated that perturbative formulas should be trustworthy to evolve $\alpha(q)$ to yet higher energies. The desired link between low and high energy has thus been constructed. We plan to extend the method to dynamical fermions in the future.